\documentclass{aa}
\usepackage{graphicx}
\begin{document}
\title{Nuclear star formation in NGC~6240}
\author{Anna Pasquali,
        \inst{1,2}
        John S. Gallagher
        \inst{3}
        \and
        Richard de Grijs
        \inst{4}}
\offprints{pasquali@phys.ethz.ch}
\institute{ESO/ST-ECF,
             Karl-Schwarzschild-Strasse 2, 85748
             Garching bei M\"unchen, Germany
          \and
             Institute of Astronomy, ETH Hoenggerberg,
             8093 Z\"urich, Switzerland\\
             \email pasquali@phys.ethz.ch
           \and
           University of Wisconsin,
           Department of Astronomy,
           475 N. Charter St., Madison WI 53706, USA\\
           \email jsg@astro.wisc.edu
           \and
           University of Sheffield,
           Department of Physics and Astronomy,
           Hicks Building, Hounsfield Road,
           Sheffield S3 7RH, UK,
           \email R.deGrijs@sheffield.ac.uk}
\date{Received; submitted}

\abstract{We have made use of archival HST BVIJH photometry to constrain the
nature of the three discrete sources, A1, A2 and B1, identified in the double
nucleus of NGC~6240. STARBURST99 models have been fitted to the observed
colours, under the assumption, first, that these sources can be treated as
star clusters (i.e. single, instantaneous episodes of star formation), and
subsequently as star-forming regions (i.e. characterised by continuous star
formation). For both scenarios, we estimate ages as young as 4 million years,
integrated masses ranging between 7 $\times$ 10$^6$ M$_{\odot}$ (B1) and
10$^9$ M$_{\odot}$ (A1) and a rate of 1 supernova per year, which, together
with the stellar winds, sustains a galactic wind of 44 M$_{\odot}$ yr$^{-1}$.
In the case of continuous star formation, a star-formation rate has been
derived for A1 as high as 270 M$_{\odot}$ yr$^{-1}$, similar to what is
observed for warm Ultraluminous Infrared Galaxies (ULIRGs) with a double
nucleus. The A1 source is characterised by a mass density of about 1200
M$_{\odot}$ pc$^{-3}$ which resembles the CO molecular mass density measured
in cold ULIRGs and the stellar density determined in ``elliptical core''
galaxies. This, together with the recent discovery of a supermassive binary
black hole in the double nucleus of NGC~6240, might indicate that the ongoing
merger could shape the galaxy into a core elliptical. 
}
\titlerunning{Nuclear star formation in NGC~6240}
\maketitle

\section{Introduction}

The importance of starbursts associated with major mergers between galaxies
became clear when IRAS revealed populations of luminous and ultraluminous
infrared galaxies (LIRGs and ULIRGs) in the nearby universe (cf. Neugebauer
et al. 1984, Soifer et al. 1984). 
Interest therefore remains high in understanding how these starbursts are
structured. While LIRGs are characterised by an infrared (IR; $5 - 500 \mu$m)
luminosity L$_{\rm IR} >$ 10$^{11}$ L$_{\odot}$, ULIRGs exceed L$_{\rm IR}
\simeq$ 10$^{12}$ L$_{\odot}$ (Sanders \& Mirabel 1996). ULIRGs contain $10^8
- 10^{9.5}$ M$_{\odot}$ of molecular gas (Evans et al. 2001) and are
typically the merger of two or more disk galaxies of nearly equal mass ($<$
3:1, Sanders 2001). Indeed, NGC 4038/39 (the Antennae), Arp 220 and NGC 7252 
are not only classical examples of LIRGs and ULIRGs, but also are key benchmarks 
in the Toomre sequence of galaxy mergers (Toomre 1977).

HST/WFPC2 and NICMOS imaging has revealed that 5\% -- 20\% of the known
ULIRGs are multiple mergers, probably descending from compact groups of
galaxies (Cui et al. 2001, Bushouse et al. 2002). Independently of their
multiplicity, ULIRGs show circumnuclear bright knots usually identified with
concentrations of stars younger than $10^7 - 10^8$ years (e.g., Zepf \&
Ashman 1993, Scoville et al. 2000, Surace et al. 2000, Farrah et al. 2001). At
optical wavelengths, ULIRGs are characterised by a nuclear spectrum which is
either Seyfert-like consistent with the presence of an AGN or QSO, LINER-like
or HII-region-like. In these last two cases, star clusters and/or HII regions
seem to be the dominant source of ionization (Veilleux et al. 1999).

NGC~6240 is one of the closest member of the class of ULIRGs, being at a
distance of $\simeq 98$ Mpc (for $H_0 = 75$ km s$^{-1}$ Mpc$^{-1}$). Its far-IR
luminosity is $(3 - 11) \times 10^{11}$ L$_{\odot}$ (Wright et al. 1984) and
it places NGC~6240 at the faint (or ``warm'') limit of the ULIRGs class.
>From the point of view of morphology, NGC~6240 is classified as a merger of
two massive disk galaxies (Zwicky et al. 1961, Fosbury \& Wall 1979, Tacconi
et al. 1999); the colours of the nuclear and circumnuclear stellar populations
indicate that the merger has possibly been ongoing for the past $\simeq$ 1 Gyr
(Genzel et al. 1998, Tecza et al. 2000). The gravitational interactions as
well as disturbances induced by star formation are responsible for large-scale
dust lanes, loops, shells and tails extending out to $\sim 30$ kpc (cf.
Pasquali et al. 2003, hereafter Paper I). The central region of the merger is
characterised by a double nucleus, with the two optically visible nuclei
separated by $1.''5 - 2''$ on average with their apparent separation
increasing at shorter wavelengths (Bryant \& Scoville 1999, Scoville et al.
2000). The H$_2$ and CO emission is found to peak in between the nuclei,
possibly in a thick disk structure, and slightly closer to the Southern
nucleus (van der Werf et al. 1993, Tacconi et al. 1999, Ohyama et al. 2000).
This off-set is believed to be the result of interactions between the
molecular gas and an expanding shell centered on the southern nucleus (Rieke
et al. 1985), or to the gravitational settling of gas into a nuclear disk as
the interaction evolves (Tacconi et al. 1999).

A remarkable feature of the NGC~6240 nuclei is the three discrete sources, A1,
A2 and B1 located in the southern ``A'' and the northern ``B'' nucleus,
respectively. They were first resolved in HST/FOC B-band images by Barbieri et
al. (1993) who measured an enhanced blue continuum emission corresponding to
these sources. In a subsequent analysis based on [OII]/[OIII]/H$\beta$ line
ratios, Rafanelli et al. (1997) suggested that B1 is a LINER, and A2 a LINER
or a HII region (cf. Veilleux et al. 1995). The authors also pointed out that the
line ratios derived for A1 are typical of a Seyfert galaxy or a
high-excitation HII region. A Seyfert-like nature would imply the presence of
an AGN in the A core. These nuclear sources are believed to drive a
large-scale outflow (Heckman et al. 1987).
 
K-band imaging of NGC~6240 reveals the presence of red supergiants in the
double nucleus (Tecza et al. 2000) from which an age between 15 and 25 Myr is
inferred for the galaxy nuclei. In the radio, 70$\%$ of the flux at 20 cm is
emitted by A and B, of which half comes from two unresolved sources and half
is diffuse emission (Colbert et al. 1994). These compact sources do not align
up with B1 and A1$+$A2, but are shifted westward and less separated, probably
because of obscuration effects. Their spectrum is quite flat ($\alpha$ = 0.6)
suggesting that their radio emission is powered by the local starburst.

Beswick et al. (2001) observed neutral hydrogen in absorption against the two
nuclei of NGC~6240 using MERLIN with an angular resolution of 0.2~arcsec. The
nuclei appear as slightly extended sources at a frequency of 5~GHz with a
resolution of 0.102$\times$0.055~arcsec and are separated by 1.52 arcsec. The
radio continuum emission is attributed to a combination of AGNs and starburst
activity. The HI absorption appears to be associated with the gas disk that is
also seen in CO. Hagiwara, Diamond, \& Miyoshi (2003) present radio maps of
the H$_2$O masers in NGC~6240, and show that the masers are concentrated in the
southern A1 nucleus, which evidently contains dense gas.

ISO data have revealed OIV emission from the nuclei which is attributed to an
AGN (Genzel et al. 1998, Lutz et al. 2003). AGN-like hard X-ray emission is
observed below 10 keV (Iwasawa \& Comastri 1998, Vignati et al. 1999).
High-resolution follow-up observations with Chandra and the HRC camera in the
$0.08 - 10$ keV band have been performed by Lira et al. (2002). In the high
resolution X-ray map, the emission peaks coincide with B1 and A1. Lira et al.
used the X-ray fluxes together with optical-to-radio luminosities measured for
the galaxy's double core and fitted the resulting spectral energy distribution
(SED) with a combination of starburst, QSO and blackbody continuum spectra. It
then turns out that the starburst component dominates the observed SED at most
wavelengths and the dust associated with the starburst is responsible for
the K-band emission.
The AGN accounts mainly for the hard X-ray emission and is
completely absorbed in the near IR, optical and soft X-ray bands. Very
recently, Komossa et al. (2003) imaged the A and B nuclei with Chandra ACIS-S
and clearly detected an AGN in each nucleus, so that the existence of a binary
system of supermassive black holes in the core of NGC~6240 has now been firmly
established.

In this paper, we aim to estimate the age, mass and reddening of the stellar
content of the nuclear sources in NGC~6240 and compare these with their
properties at radio and X-ray wavelengths, in order to: {\it i)} discuss the
star formation mode ongoing in the core of such a gas-rich and dynamically
extreme system, and {\it ii)} possibly shed some light on the nature of the
end-product of the merging process at work in NGC~6240.

\section{Data processing}
\subsection{Observations}

NGC~6240 was observed with HST at several wavelengths. Here, we have collected
the data acquired with WFPC2 as part of the GO proposal 6430 (PI van der
Marel, cf. Gerssen et al. 2001) and NICMOS/NIC2 for the GO proposals 7219 (PI
Scoville) and 7882 (PI van der Werf). The image rootnames with their
corresponding filters and exposure times are listed in Table 1.

\begin{table}
\caption[]{The log of NGC~6240 images taken with HST/WFPC2 and NICMOS.}
\begin{tabular}{ccr}
\hline
Dataset    & Filter          & Exp. Time in s\\
\hline
\noalign{\smallskip}
u4ge010..  & F450W - broad B  & 3 $\times$ 700.00 \\
u4ge010..  & F547M -- medium V & 400.00 $+$ 800.00 \\
u4ge010..  & F814W -- I        & 3 $\times$ 400.00 \\
           &       & \\
n48h09w..  & F110W -- J        & 4 $\times$ 39.95 \\
n48h09...  & F160W -- H        & 4 $\times$ 47.95 \\
n48h09x..  & F222M -- K short  & 4 $\times$ 55.94 \\
\hline
\end{tabular}
\end{table}

We retrieved the WFPC2 and NICMOS images already pipeline-processed, i.e.
corrected for bias, dark current and flat field. We registered the WFPC2
frames to the same spatial grid of the images taken with the F450W filter, by
simply measuring the position of several point sources in common and their
relative shifts in (X;Y) among the available datasets. The (X;Y) shifts were
then applied with the IRAF routine IMSHIFT. Once aligned, the images acquired
with the same filter were combined with the STSDAS task CRREJ to clean them
from cosmic rays and have them median-combined to improve their S/N ratio.
\par\noindent The NICMOS pipeline-reduced images were aligned to the frames
acquired with the F110W filter with the same technique as above. Since they
were obtained with Camera 2 (FOV of 22$'' \times$ 22$''$ at 0.075$''$/pix)
they essentially overlap with the central area of the PC images containing the
galactic double nucleus.

\subsection{Multiband photometry}  

The high spatial resolution of the Planetary Camera has made possible to
detect three discrete sources in the nuclei of NGC~6240 which were labelled A
and B by Barbieri et al. (1993). On the basis of their coordinates, our
sources have counterparts in Barbieri et al.'s FOC images, namely A1, A2 and
B1, and in the Chandra images at $0.08 - 10$ keV obtained by Lira et al. (2002).
Figure 1 shows the F814W and F160W images of these nuclear sources, where the
NIC2 image in H has been scaled to the resolution of the PC I image.

\begin{figure}
\centering
\includegraphics[width=6.5cm]{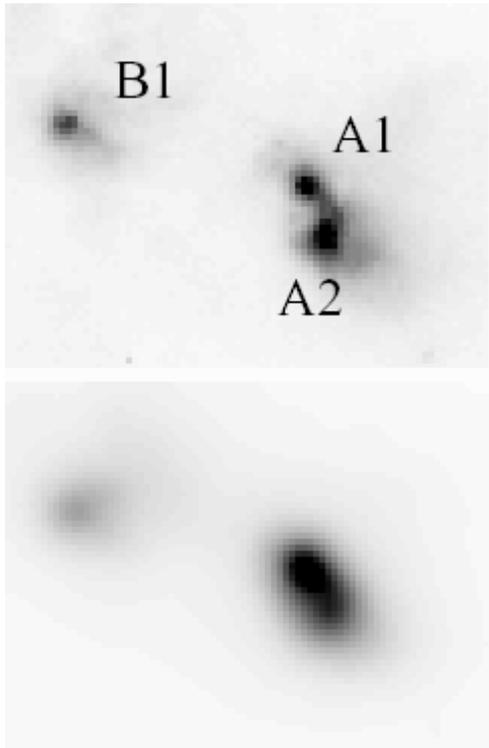}
\caption{The nuclear sources in NGC~6240, A1, A2 and B1, as they appear in the
PC I image (top) and in the NIC2 H image (bottom). The latter has been scaled
to the PC angular resolution.}
\end{figure}

\begin{table*}
\caption[]{The observed magnitudes of the discrete sources in NGC~6240
nuclei.}
\begin{tabular}{cccccc}
\hline
Cluster&B&V&I&J&H \\
\hline
\noalign{\smallskip}
A1 &23.27 $\pm$ 0.05 &20.82 $\pm$ 0.04 &18.73 $\pm$ 0.02 &17.49 $\pm$ 0.02 &16.96 $\pm$ 0.02\\
A2 &21.59 $\pm$ 0.03 &19.77 $\pm$ 0.04 &18.56 $\pm$ 0.05 &17.73 $\pm$ 0.03 &17.33 $\pm$ 0.03\\
B1 &21.19 $\pm$ 0.01 &19.84 $\pm$ 0.01 &19.24 $\pm$ 0.03 &18.66 $\pm$ 0.04 &18.49 $\pm$ 0.04\\
\hline
\end{tabular}
\end{table*}

Their positions were measured with the IRAF routine IMEXAMINE, and were used
as the input coordinates for PHOT in DAOPHOT. Photometry of the nuclear
sources was performed assuming an aperture radius of 3 pixels in the PC
images. A median, local  background of diffuse emission  underlying the 
nuclear sources was computed over an annulus of 4-to-6 pixels.

Aperture fluxes were corrected for the camera's charge transfer (in)efficiency
(CTE) following Whitmore \& Heyer (1998). Aperture corrections to the standard
0$''$.5 aperture were determined from two bright and point-like isolated
objects in the PC, and subsequently applied to the measured fluxes. Only at
this stage, fluxes were first transformed into the WFPC2 synthetic magnitude
system and subsequently into the Johnson magnitude system using the zero
points and the colour equations listed in Holtzman et al. (1995).

Because it becomes very difficult to clearly identify the A1, A2 and B1
sources in the K band, where emission from dust dominates the stellar emission
from the nuclei (cf. Gerssen et al. 2003), we performed aperture photometry
only on the F110W and F160W images. We set the aperture radius to 2 pixels and
measured a median sky in an annulus of 5-to-8 pixels. Subsequently,
background-corrected aperture fluxes were scaled first to those corresponding
to the fluxes in a 0$''$.5 radius aperture and then to the flux of an infinite
radius aperture following the prescription given in the NICMOS Photometry
Cookbook (cf. http:/www.stsci.edu/hst/nicmos/performance/\\
photometry/cookbook.html).

The BVIJH magnitudes of the nuclear sources and their uncertainties are listed
in Table 2.

\section{The stellar populations in the nuclear sources}

The A1, A2 and B1 sources have a mean FWHM of 5.2, 9.6 and 4.2 pixels
respectively, a factor of 3 to 6 larger than the PC stellar PSF, therefore
they are resolved. From their FWHM we can derive estimates for their tidal radii,
which vary between 47 and 108 pc. Such values have also been estimated for
some of the younger and intermediate-age star clusters in the Antennae
(Whitmore et al. 1999): knot S and cluster $\#$430 have a tidal radius of 450
pc and $>$ 73 pc, respectively, while the intermediate-age cluster $\#$225 has
a tidal radius of 50 pc. In comparison, all the globular clusters in M31 are
characterized by R$_t <$ 100 pc (Cohen \& Freeman 1991, Grillmair et al.
1996), and only two Galactic globular clusters have R$_t >$ 200 pc (Djorgovski
1993). In addition, young and massive star clusters are quite often found in
the nuclei of late-type spiral and barred spiral galaxies (cf. Sect. 5). On
the other hand, the nuclear sources in NGC~6240 possibly show spectra typical
of HII regions (Veilleux et al. 1995, Rafanelli et al. 1997) and may well be
extended clumps of star formation. This option would also be more consistent
with the findings of an underlying, $\simeq$ 1 Gyr-old stellar population
(Genzel et al. 1998, Tecza et al. 2000), so that the light of the newly
born OB stars is now dominating the colours of the nuclear sources.

We have fitted the colours observed for the nuclear sources [(B-V), (V-I),
(V-J) and (V-H) corrected for the Galactic E(B-V) = 0.076 mag (Schlegel
et al. 1998)] to STARBURST99 (Leitherer et al. 1999) tracks, which were
computed: 
\par\noindent 
{\it i)} for instantaneous star formation in a star cluster with a mass of
10$^5$ M$_{\odot}$ and solar metallicity (cf. Sect. 3.1);
\par\noindent 
{\it ii)} for continuous star formation with a rate (SFR) of 10 M$_{\odot}$
yr$^{-1}$ and solar metallicity (cf. Sect. 3.2).
\par\noindent 
Indeed, colours depend mainly on reddening and age once the metallicity is
fixed. In massive clusters stochastic effects in the stellar population 
are minmized and so the colors become insensitive to the total mass so 
long as the stellar mass function is fixed (Lan\c{c}on \& Mouchine 2000,
Cervi\~no et al. 2002).
No metallicity has so far been measured for NGC~6240, but
McCall et al. (1985) predicted a metallicity of about 20$\%$ solar. The
computed STARBURST99 tracks also assume a Salpeter IMF from 0.1 M$_{\odot}$ to
100 M$_{\odot}$, a (Type II) supernova cut-off mass of 8 M$_{\odot}$, standard
stellar mass loss and theoretical wind models.
 
The theoretical colours at each time mesh point were reddened according to
Calzetti's (2001) extinction law, and a $\chi^2_{\rm tot}$ value was computed
as the sum of the differences between the reddened theoretical and the
observed colours, weighted by the observational uncertainties. For each time
mesh-point, the reddening was varied from E(B-V) = 0.0 to 5.0 mag with a step
of 0.01 mag and a set of reddening, absolute V magnitude and colours have been
saved corresponding to the minimum $\chi^2_{\rm tot}$ derived from the fit to
the observed colours. The lowest value among the $\chi^2_{\rm tot}$ minima
($\chi^2_{\rm min}$) obtained along the age sequence of the evolutionary 
track has been selected
together with its relative set of best-fitting reddening, colours and absolute
magnitudes. The latter represent our best-fitting model SED to each of the
three discrete sources. In addition, the reddened theoretical apparent V
magnitude was scaled to the observed one and the initial track mass or SFR was
multiplied by this scaling factor in order to estimate either the integrated
mass or the total SFR of the discrete sources.

In order to assign a confidence interval to age, mass and reddening, we have
extracted from the previous fits all the (reddening, age, absolute magnitudes,
mass) solutions associated with a $\chi^2_{\rm tot} \le$ 1.5 times the 
$\chi^2_{\rm MIN}$ value. These are selected to obtain the range of
acceptable suboptimal fits to the photometry. We have used
these intervals to construct the probability functions of age, mass and
reddening for A1, A2 and B1 in the same way as done in Paper I for the
clusters in the main body and tails of NGC~6240. In summary, we combined the
ranges spanned by the fits to age, reddening and mass into histograms, where the
value of each bin is the number of acceptable solutions. Each bin was then
normalized by the total number of model solutions for the nuclei of NGC~6240,
so that a probability function is obtained for the age, mass and
reddening of the sources. A word of caution applies here, whereby a photometry
measured through larger apertures might give somewhat higher masses and SFRs
and might negligibly affect ages. 

\subsection{The cluster hypothesis}

\begin{figure*}
\centering
\includegraphics[width=15cm]{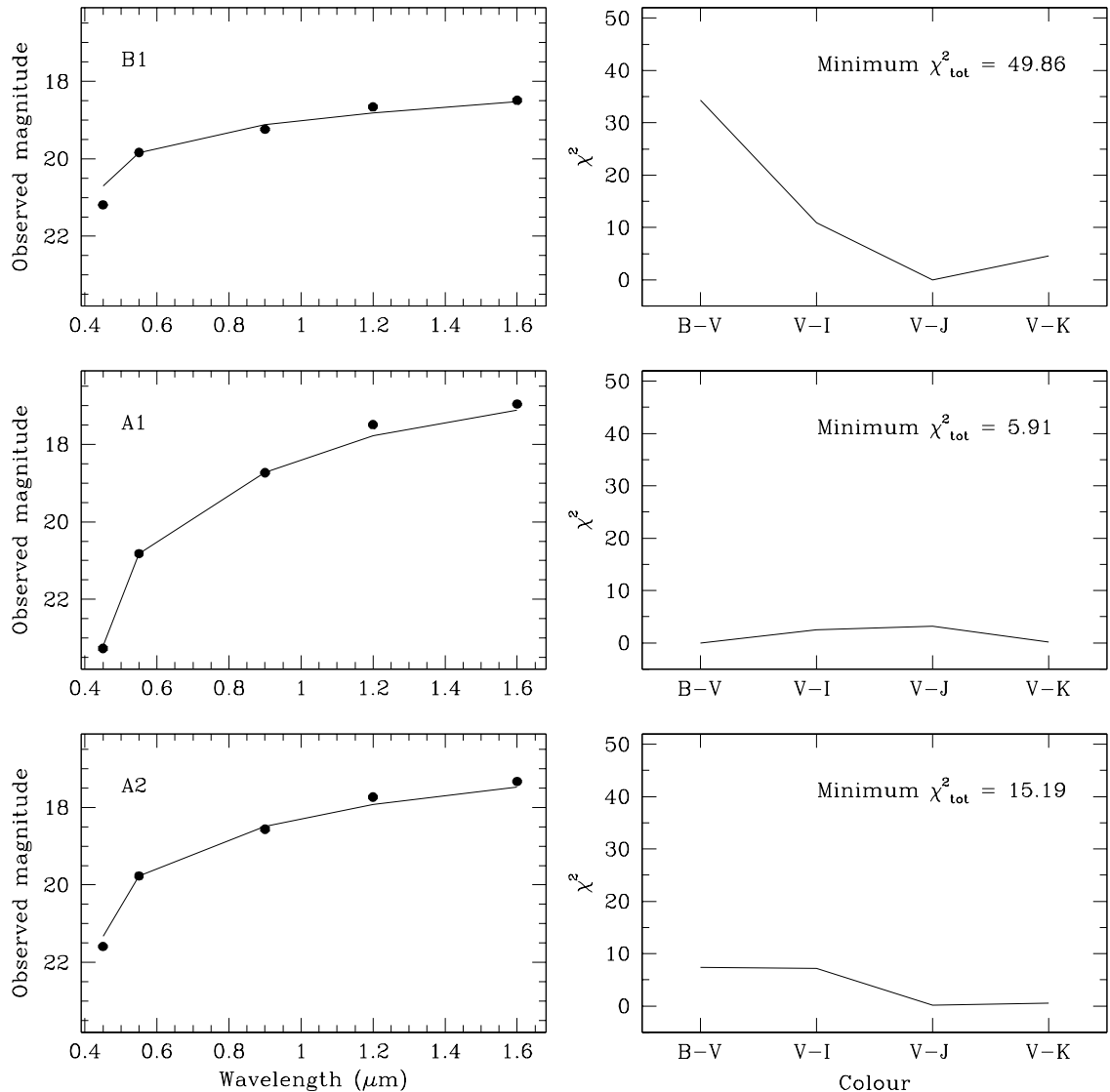}
\caption{The best-fitting spectral energy distributions (solid lines) obtained
for the discrete sources A1, A2 and B1 in the nuclei of NGC~6240 with
superimposed the observed, apparent magnitudes. The assumption here is that
A1, A2 and B1 are star clusters. In the right panels, the $\chi^2$ value is
plotted for each source and colour for minimum $\chi^2_{\rm tot}$.}
\end{figure*}

The best-fitting SEDs obtained under the assumption that the nuclear sources
in NGC~6240 are star clusters are plotted, as solid lines, in the left-hand
panels of Fig. 2. Superposed are the sources observed, apparent
magnitudes, represented by filled dots. The observational uncertainties (cf.
Table 2) are smaller than the symbol sizes. The $\chi^2$ obtained for each
colour for minimum $\chi^2_{\rm tot}$ is plotted in the right-hand panels of
Fig. 2 with the aim of showing which colours agree better with the
evolutionary tracks and which are discrepant. The lowest $\chi^2_{\rm tot}$
($\simeq$ 6) occurs for the A1 source, whose theoretical SED agrees with the
observations to within the photometric errors (except for the J band). The B1
and A2 best-fitting SEDs deviate from the observed B band by 0.3 and 0.5 mag
respectively, and this discrepancy could in part be due to contamination 
by nebular
emission (see, e.g., Anders \& Fritze-von Alvensleben 2003 for possible
contaminants and their effects on broad-band fluxes as a function of age of
the dominant stellar population). We have estimated a line-contribution
of $\sim$ 0.1 mag in the F450W filter from the integrated central spectrum
published by Fosbury \& Wall (1979) and $\sim$ 0.07 - 0.05 mag in the F110W and
F160W filters from the spectra in Simpson et al. (1996).
Uncertainties in the colour equations used
to transform WFPC2 magnitudes to the Johnson system may also contribute. 
The best-fitting age, reddening and mass of
A1, A2 and B1 are listed in Table 3. Adopting Calzetti's (2001) extinction
law, we estimate for the double nucleus of NGC~6240 an A$_V$ extinction between
4 and 9 mag, in agreement with the value 3 $\le$ A$_V \le$ 5 initially derived
by Fosbury \& Wall (1979) and DePoy et al. (1986). The estimate given by Lutz
et al. (2003) is larger, A$_V \simeq 10 - 15$ mag, which has been derived from
the flux ratio of the Br$\gamma$ to the [NeII] 12.81$\mu$m lines. The authors
themselves consider their estimated A$_V$ very uncertain because of the
difficulty to measure the Br$\gamma$ line of the underlying gas against the
stellar component.

\begin{table}
\caption[]{The best-fitting age, reddening and mass for A1, A2 and B1 computed
under the assumption that these sources are stellar clusters.}
\begin{tabular}{cccc}
\hline
Source  & Age (Myr)   & E(B-V) (mag) & Mass (M$_{\odot}$) \\
\hline
\noalign{\smallskip}
A1   & 4.4 $\pm$ 0.05 & 2.3 $\pm$ 0.05 & 1.2 ($\pm$ 0.2) $\times$ 10$^9$\\
A2   & 3.9 $\pm$ 0.10 & 1.7 $\pm$ 0.02 & 4.2 ($\pm$ 0.5) $\times$ 10$^8$\\
B1   & 4.0 $\pm$ 0.05 & 1.0 $\pm$ 0.02 & 2.9 ($\pm$ 0.2) $\times$ 10$^7$\\
\hline
\end{tabular}
\end{table}

The probability functions (for the sources' age, mass and reddening) are
plotted in Fig. 3 as dashed histograms, together with those derived in Paper
I for the clusters in the main body and tails of NGC~6240. A caveat applies
here, as discussed extensively in Paper I: the histograms in Fig. 3 simply
give the most probable value(s) for the clusters' age, mass and reddening, but
do not represent the true cluster parameters. Indeed, when the mass ranges
spanned by the clusters are combined in a probability function, their width
(i.e. the error bar on the mass value most consistent with the data for each
cluster) has the effect of smearing any intrinsic relationship between
apparent luminosity and cluster mass, and the probability function may show a
peak. In addition to this, evolutionary effects, such as those associated with
a non-coeval population of star clusters, also broadens power law luminosity
functions (cf. Meurer 1995, Fritze-von Alvensleben 1998, 1999, de Grijs et al.
2001, 2003a). Moreover, luminosity selection effects (i.e. limiting magnitude
of $\simeq$ 25 mag) prevent the detection of clusters at any age that are less
massive than 4 $\times$ 10$^4$ M$_{\odot}$ and allow only ever more massive
clusters to be revealed at gradually older ages. Therefore, the mass
distribution of the NGC~6240 clusters (middle panel in Fig. 3) is heavily
compromised by the limiting magnitudes for masses lower than 4 $\times$ 10$^4$
M$_{\odot}$, even among the very young star clusters.

Because the fits for the nuclear sources are based on four independent
colours, instead of two as for the other NGC~6240 star clusters (cf. Paper I),
the confidence intervals for A1, A2 and B1 are well determined and narrow:
they do not overlap, except in age. The discrete sources in the nuclei of NGC
6240 certainly stand out because of their young ages with respect to the
overall cluster population, but, while B1 and A2 fall on the tail of the
cluster mass and reddening distributions, A1 shows extreme values for both of
these parameters. It is interesting to note the increase in E(B-V) from B1 in
the northern nucleus to A1 and A2 in the southern one which is consistent
with the 3D geometry of the double nucleus as derived by Ohyama et al. (2003).

\begin{figure}
\centering
\includegraphics[width=8.5cm]{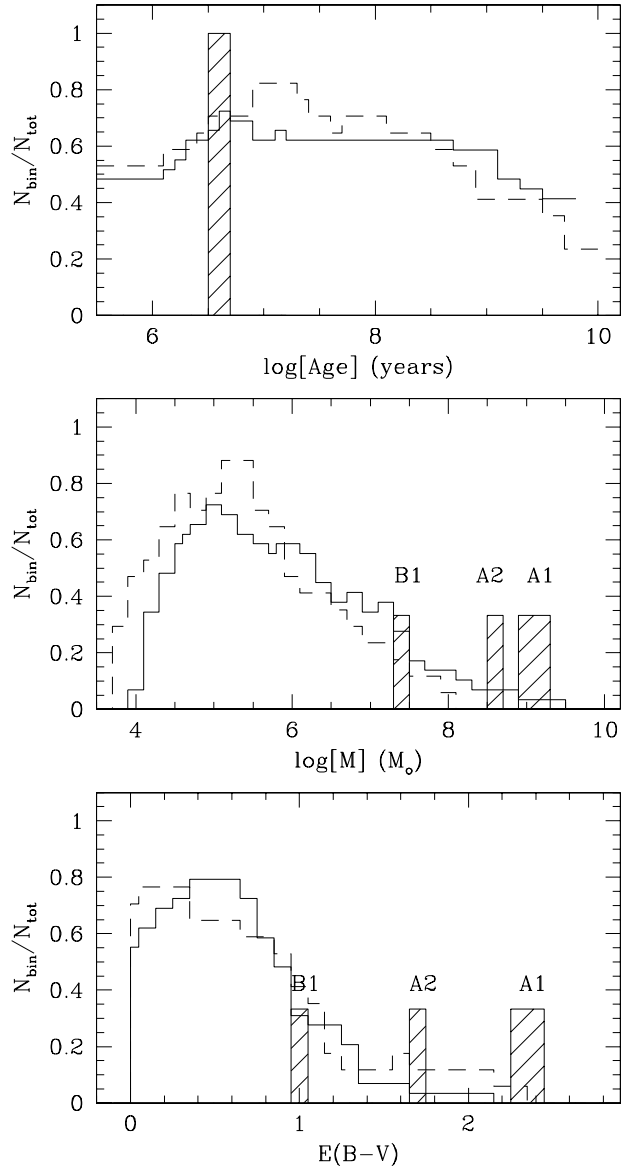}
\caption{The probability distributions of age and mass in logarithmic units,
and reddening on a linear scale. The solid line represents the probability
distribution for the population of star clusters detected in the NGC~6240 main
body, the dashed line traces the clusters identified in the galaxy's tails
(cf. Paper I). The dashed bins show the most probable values from fits to the
A1, A2 and B1 sources.}
\end{figure}

\subsection{The hypothesis of continuous star formation}

Only a very small number of compact coeval star-forming events with masses
like those inferred for the nuclei of NGC~6240 in the previous Section have
been observed (e.g., NGC 7252 cluster W3: Schweizer \& Seitzer 1998, Maraston
et al. 2001; NGC 6745: de Grijs et al. 2003b); spectroscopic mass confirmation
has only been obtained for cluster W3 in NGC 7252 to date. The presence of an
AGN within both nuclei also points to complex evolutionary histories where we
expect circumnuclear stars to span a range in age. We further hypothesize that
the circumnuclear SFR is likely to have increased during the ongoing merger in
NGC~6240. Therefore we have to consider a continuous star formation scenario,
such as described by STARBURST99 models with a SFR of 10 M$_{\odot}$ yr$^{-1}$
and solar metallicity. The resulting best-fitting model
SED is shown
in Fig. 4 for each source, where we also indicate the $\chi^2$ associated
with the observed colours for minimum $\chi^2_{\rm tot}$, and the best-fitting
properties of the three nuclear regions are listed in Table 4. Similarly as
for the cluster scenario, we determine for the galaxy double core an A$_V$
extinction ranging from 3 to 9 mag.

\begin{table*}
\caption[]{The best-fitting age, reddening, SFR and integrated mass for A1, A2
and B1 computed under the assumption that these sources are star-forming
regions and governed by continuous star formation.}
\begin{tabular}{ccccc}
\hline
Source  & Age (Myr)   & E(B-V) (mag) & SFR (M$_{\odot}$ yr$^{-1}$) & Integrated Mass (M$_{\odot}$) \\
\hline
\noalign{\smallskip}
A1      &4.4 $\pm$ 0.9 & 2.4 $\pm$ 0.07 & 266 $\pm$ 109 & 1.3 ($\pm$ 1.0) $\times$ 10$^9$ \\
A2      &4.3 $\pm$ 0.4 & 1.5 $\pm$ 0.05 & 21.5 $\pm$ 5.7& 9.6 ($\pm$ 3.4) $\times$ 10$^7$ \\
B1      &4.3 $\pm$ 0.7 & 0.8 $\pm$ 0.08 & 1.5 $\pm$ 0.6 & 7.2 ($\pm$ 3.7) $\times$ 10$^6$ \\
\hline
\end{tabular}
\end{table*}

The minimum $\chi^2_{\rm tot}$ is particularly large for B1 and A2 as it is
dominated by the large discrepancy in the (B-V) colour (0.7 and 0.5 mag,
respectively). As already suggested, this deviation could in part be
produced by contamination from nebular emission in the B band; 
uncertainties in the colour equations used to transform WFPC2 magnitudes to
the Johnson system may also contribute. In the case of B1 and A2, the minimum
$\chi^2_{\rm tot}$ is a factor of 2 worse than obtained under the assumption
that these sources are star clusters. The SEDs derived for A1 for the two
scenarios are similar ($\chi^2_{\rm tot}$ of 5.91 for a cluster SED against
6.56 for a continuous star-forming region). The difference in (B-V) and (V-I)
between the evolutionary tracks employed so far (with a fixed cluster mass of
10$^5$ M$_{\odot}$ and for a SFR of 10 M$_{\odot}$ yr$^{-1}$) are within the
observational errors for ages of few million years. Therefore, given the
observed colours of A1, A2 and B1, we are not able to discriminate between a
cluster and a sharp SBST model.

Probability functions have also also derived and are plotted as dashed
histograms in Fig. 5, together with the probability functions computed for
the NGC~6240 clusters (from Paper I).

Once again, the discrete sources in the nuclei of NGC~6240 are among the
youngest stellar systems in the galaxy, with ages between 3.5 to 5.4 $\times$
10$^6$ years, thus implying that young stars dominate the circumnuclear
regions in optical. The integrated mass of A1, A2 and B1 overlaps with the
high-mass tail of the overall NGC~6240 cluster distribution, with A1 showing
an extreme value once again. This same trend can be recognized in the
reddening histogram.

\begin{figure*}
\centering
\includegraphics[width=15cm]{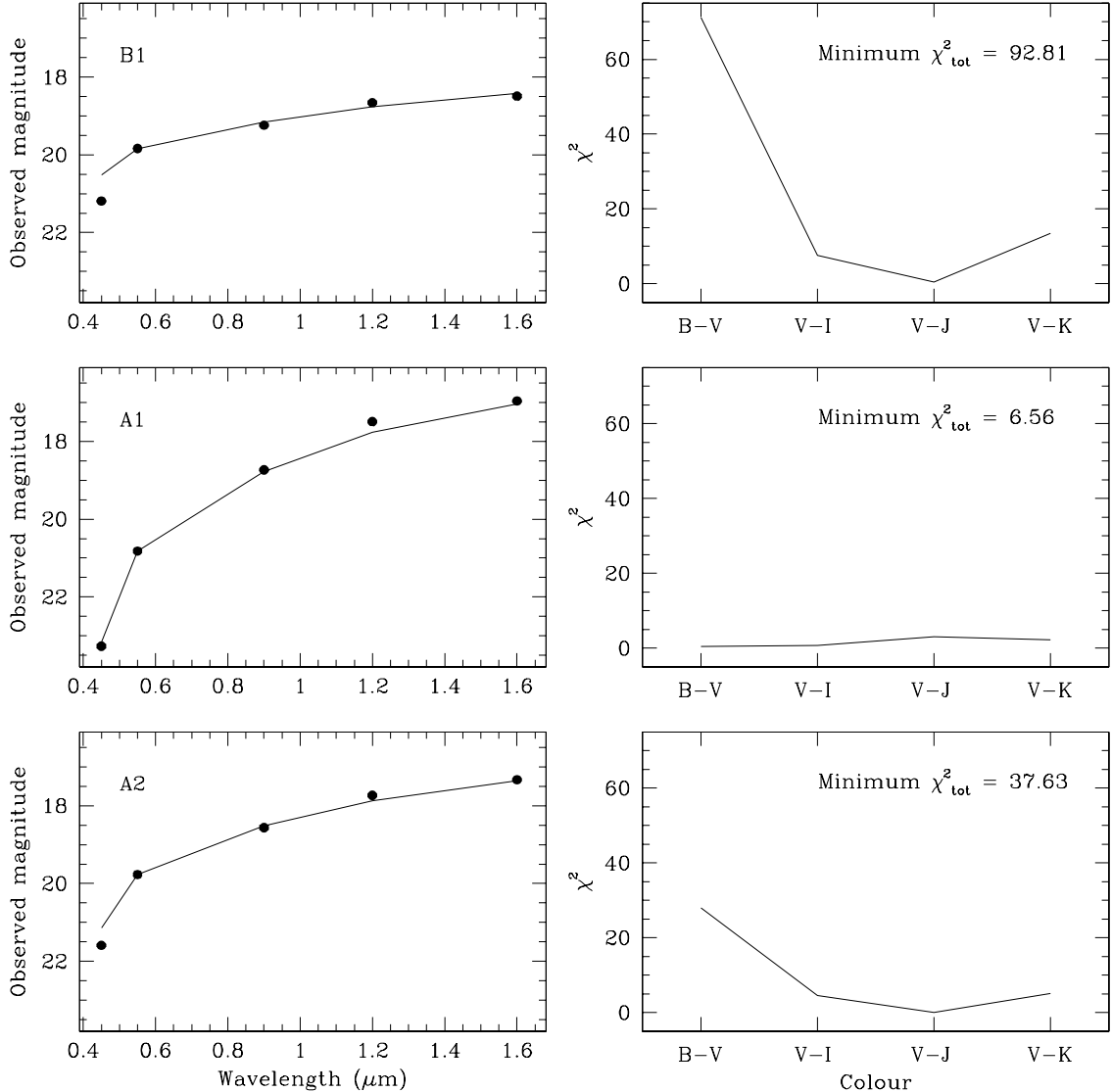}
\caption{As Fig. 2, but now assuming a continuous star formation scenario.}
\end{figure*}

\subsection{Dependences on metallicity}
To test the robustness of our results against metallicity, we have
repeated the above analysis using STARBURST99 tracks computed 
under the same assumptions, but with the metallicity set to
Z = 0.008 (LMC-like). Tables 5 and 6 
summarize the best-fitting parameters obtained for the cluster
and the continuous star formation scenarios, respectively.

\begin{table*}
\caption[]{The best-fitting age, reddening and mass for A1, A2 and B1 computed
under the assumption that these sources are stellar clusters with an
LMC-like metallicity, Z = 0.008.}
\begin{tabular}{ccccc}
\hline
Source  & Age (Myr)   & E(B-V) (mag) & Mass (M$_{\odot}$) & $\chi^2_{\rm tot}$\\
\hline
\noalign{\smallskip}
A1   & 4.7 $\pm$ 2.0 & 2.4 $\pm$ 0.2 & 2.6 ($\pm$ 1.4) $\times$ 10$^9$ & 6.39\\
A2   & 4.3 $\pm$ 1.5 & 1.6 $\pm$ 0.2 & 4.2 ($\pm$ 2.8) $\times$ 10$^8$ & 18.87\\
B1   & 4.6 $\pm$ 1.6 & 0.9 $\pm$ 0.2 & 2.0 ($\pm$ 1.2) $\times$ 10$^7$ & 57.05\\
\hline
\end{tabular}
\end{table*}

\begin{table*}
\caption[]{The best-fitting age, reddening, SFR and integrated mass for A1, A2
and B1 computed under the assumption that these sources are star-forming
regions and governed by continuous star formation. Metallicity is now set to
Z = 0.008.}
\begin{tabular}{cccccc}
\hline
Source  & Age (Myr)   & E(B-V) (mag) & SFR (M$_{\odot}$ yr$^{-1}$) & Integrated Mass (M$_{\odot}$) 
&  $\chi^2_{\rm tot}$\\
\hline
\noalign{\smallskip}
A1      &6.1 $\pm$ 2.5 & 2.4 $\pm$ 0.08 & 412 $\pm$ 219 & 2.5 ($\pm$ 1.8) $\times$ 10$^9$ & 5.86\\
A2      &6.2 $\pm$ 2.7 & 1.5 $\pm$ 0.13 & 33.4 $\pm$ 21.6& 2.1 ($\pm$ 1.7) $\times$ 10$^8$ & 33.64\\
B1      &6.5 $\pm$ 3.6 & 0.8 $\pm$ 0.21 & 2.8 $\pm$ 2.3 & 1.8 ($\pm$ 1.6) $\times$ 10$^7$ & 85.73\\
\hline
\end{tabular}
\end{table*}

\par\noindent
We conclude that the properties derived for a solar- and a LMC-like
metallicity are well within the uncertainties of the best fits for each of the
scenarios modelled. A better overall agreement between the solar and LMC-like
metallicity cases is achieved for the cluster hypothesis. In the case of
continuous star formation, the sources' ages and star-formation rates are
systematically larger for Z = 0.008 than for Z = Z$_{\odot}$. Although the
$\chi^2_{\rm tot}$ values of the best fits are comparable, an LMC-like
metallicity introduces larger uncertainties on the parameters than a solar
chemical composition. Hereafter, we will adopt the results of the evolutionary
tracks with Z = Z$_{\odot}$. 

\section{Discussion}

\subsection{Nuclear stellar populations}

A starbust can be modeled either as a single and instantaneous event of star
formation, or as an act of continuous star formation extended over a short
period of time. We have computed a STARBURST99 model for both scenarios, and
assuming a total cluster mass of 10$^5$ M$_{\odot}$ in the case of a single
stellar burst and a SFR of 10 M$_{\odot}$ yr$^{-1}$ for the continuous star
formation hypothesis. We have fitted the photometry measured for the multiple
nuclei of NGC~6240 A1, A2 and B1 to the synthetized populations, hoping that
the observed colours could discriminate between the two scenarios. Within the
observational uncertainties, the results almost overlap and do not support
either of the two models more strongly than the other.

\subsubsection{Ages}

Both models indicate that the optical SEDs of A1, A2 and B1 are dominated by
light from stellar populations with ages of $\simeq$ 4 Myr and
moderate-to-heavy obscuration. As already discussed by Lira et al. (2002),
these optical SEDs are not uniquely tied to the presence of the AGN in either
B1 or A1.

We notice that all three nuclear condensations in NGC~6240 resemble other
Starburst/AGN systems, where large contributions to the light come from young
stellar populations. Indeed, Gonz\'alez Delgado et al. (2001) found that young
and intermediate-age stars are present in the nuclear and circumnuclear
regions of 45$\%$ of Seyfert 2 galaxies. The contribution of young and
intermediate-age stars to the nuclear stellar population in Seyfert 2 galaxies
can be as high as 50$\%$ and 40$\%$ respectively (e.g. NGC~5135 and NGC~7130;
Gonz\'alez Delgado et al. 2001).

\subsubsection{Stellar Mass Estimates and Star Formation Rates}

Under the assumption of a single and instantaneous burst of star formation, we
have derived stellar masses of about 1 $\times$ 10$^9$, 4 $\times$ 10$^8$
M$_{\odot}$ and 3 $\times$ 10$^7$ M$_{\odot}$ for A1, A2 and B1, respectively.
If the cluster hypothesis were correct, these nuclei would be the most massive
known, exceeding by a factor $5 - 10$ the most massive globular clusters. In
particular, B1 would be the counterpart of the W3 cluster in NGC 7252 (at a
distance of $\simeq$ 15$''$ from the galactic centre; Whitmore et al. 1993)
for which Maraston et al. (2001) estimate a total mass of $\sim$ 3.7 $\times$
10$^7$ M$_{\odot}$ and an age of about 300 Myr, using high-resolution
spectroscopy. Another massive cluster has been detected in NGC 1316 with a
mass of about 1.4 $\times$ 10$^7$ M$_{\odot}$ and an age of 3 Gyr (Goudfrooij
et al. 2001). In addition, de Grijs et al. (2003b) have extended
the cluster mass function in the interacting galaxy NGC 6745 up to masses of
several $\times 10^8$ M$_{\odot}$, although their mass estimates are thus far
only based on multi-passband HST photometry. The masses of A1, A2 and B1,
combined with the clusters mass distribution in the outer (tails) and inner
regions of the galaxy (cf. Paper I), would point to a radial gradient in the
cluster mass, whereby less massive clusters would preferentially form far from
the galactic centre and more massive clusters closer. In the assumption that
the dynamical timescale of the merger is larger than the age of the nuclear
sources, we would expect to find more massive clusters in the galactic nuclei
not as the result of mass segregation, but as a consequence of larger
amounts of gas funneled towards the galactic centre by the merger dynamics.
Indeed, the central location of such
massive clusters as A1, A2 and B1 is particularly interesting. So far, there
have been a number of detections of a single massive cluster in the centre of
a galaxy. For example, Carollo et al. (1997) resolved compact sources in the
nuclei of 18 spiral galaxies (with an exponential bulge) in a total sample of
35 galaxies, for which a mass of $10^6 - 10^7$ M$_{\odot}$ and an age $\ge$
1 Gyr are inferred (Carollo et al. 2001). More recently, B\"oker et al. (2002)
identified a compact cluster at the centre of 59 late-type spiral galaxies in
a sample of 77 galaxies. For example, the cluster at the centre of NGC 4449
was dated between 6 and 10 Myr old and as massive as 4 $\times$ 10$^5$
M$_{\odot}$ (B\"oker et al. 2001). Colina et al. (2002) identified a cluster
of 4 Myr old and as massive as 10$^5$ M$_{\odot}$ in the nucleus of NGC 4303,
which is a barred spiral and thought to be a low-luminosity AGN. By analogy to
what is observed for NGC~6240 (cf. Lira et al. 2002), this nuclear cluster
dominates the optical emission of the galaxy core. Additional clusters,
although slightly younger ($3 - 3.5$ Myr) and less massive ($7 - 8 \times
10^3$ M$_{\odot}$), have been detected in the circumnuclear regions of NGC
4303 (Colina et al. 2002).

\begin{figure}
\centering
\includegraphics[width=8.5cm]{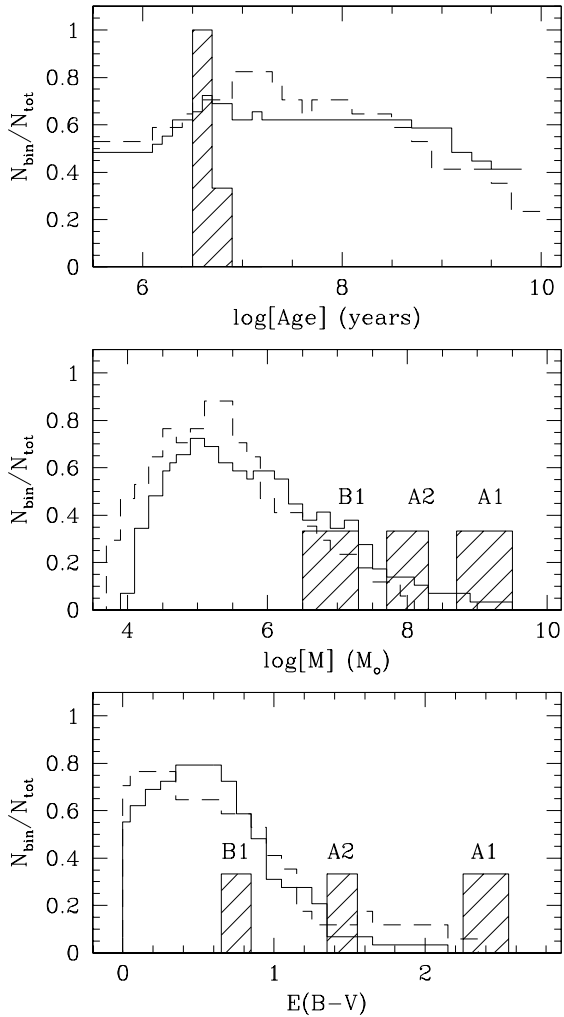}
\caption{As Fig. 3, but under the assumption that A1, A2 and B1 are
star-forming regions.}
\end{figure}

Assuming continuous star formation, we obtained for A1, A2 and B1 a SFR of
266, 21.5 and 1.5 M$_{\odot}$ yr$^{-1}$, respectively, which gives an
integrated mass of about 1 $\times$ 10$^9$, 1 $\times$ 10$^8$ and 7 $\times$
10$^6$ M$_{\odot}$ respectively. Star formation appears to be by far more
active in the southern nucleus, A, which is consistent with the southern
nucleus being brighter at radio and X-ray wavelengths, in H$_2$, CO and H$_2$O
maser (Hagiwara et al. 2003) emission. This might point to a larger reservoir
of gas in the southern nucleus compared to the northern on (see Sect. 4.2 for a
discussion on the relation between the star formation history and gas supply
in NGC 6240). In particular, the SFR estimated for A1 in the optical is of the
same order as the SFR derived from the integrated IR luminosity of NGC~6240
($\simeq$ 140 M$_{\odot}$ yr$^{-1}$, Heckman et al. 1990), thus indicating
that the nuclear sources dominate the galaxy's IR emission and star formation
activity. For these integrated masses and for the cluster masses derived
above, we estimate a total bolometric luminosity of $\sim$ 10$^{12}$
L$_{\odot}$. Lutz et al. (2003) have computed L$_{\rm bol} \simeq$ 5 $\times$
10$^{11}$ L$_{\odot}$ from comparison with starburst template spectra in the
[NeIII] 15.55$\mu$m $/$ [NeII] 12.82$\mu$m, with the caveat that this may be
an underestimate since NGC~6240 is more obscured than the template starbursts.

The SFR in the double nucleus of NGC~6240 has also been measured at radio
wavelengths by Beswick et al. (2001). They have derived, from the 1.4 GHz
luminosity, a SFR of about 83 M$_{\odot}$ yr$^{-1}$ adopting a Salpeter IMF
and assuming that stars initially more massive than 8 M$_{\odot}$ become
supernovae (SNe) with significant radio emission. This value is a factor of 2
to 5 lower than what we have estimated for A1, using BVIJ photometry and 
STARBURST99 models based on a Salpeter IMF with only stars more massive than 8
M$_{\odot}$ exploding as type II SNe. This discrepancy could have a variety 
of possible sources, including
the calibration of the techniques, possible impact of rapid evolution in
NGC~6240, and the offset between regions that are bright in the optical
and the radio, as evidenced by the radio centres being shifted westward
and having a smaller projected separation than the optical. This smaller
distance 
optical light and radio wavelengths, with the radio centres being shifted
westward and less separated. This smaller distance is probably an effect
of extinction increasing in between A1 and B1 (cf. the 3D structure of the
double nucleus as modeled by Ohyama et al. 2003). Therefore, eventual 
star-forming regions between A1 and B1 would be hidden from detection at 
optical wavelengths and become visible only at radio wavelengths. 
As a caveat, the radio SFR could in principle be associated with
star-forming regions different from A1, A2 and B1, which are not detected at
optical wavelengths because of their high extinction.
However, despite these
uncertainties, the point remains that NGC~6240 has a prodigious SFR in its
double nuclei.

In terms of SFRs and ages, the nuclear sources in NGC~6240 closely resemble
those detected in Arp~299. The latter appear to be $4 - 7$ Myr old, with SFRs
between 40 and 140 M$_{\odot}$ yr$^{-1}$ and integrated masses in the range
$(0.3 - 7) \times 10^8$ M$_{\odot}$ (i.e. 1 to 2 orders of magnitude more
massive than star clusters and HII regions in Arp~299, Alonso-Herrero et al.
2000). According to Ballo et al. (2003), Arp~299 harbours an active AGN in
both nuclei and would be, with NGC~6240, the only merging system with two
AGNs. Although the statistics are very poor, it would seem that the more
powerful AGNs may be coupled to larger nuclear SFRs and larger total nuclear
masses in newly born stars ($\sim$ 4 Myr old), as already pointed out by
Gonz\'alez Delgado et al. (1998).

\subsubsection{The High Energy Budget}

Irrespective of the scenario adopted (star clusters or continuous star
formation), the STARBURST99 code computes, for the inferred ages and total
masses of A1, A2 and B1 a SN rate of about 1 SN yr$^{-1}$, in agreement with
what Colbert et al. (1994) and Beswick et al. (2001) estimated from the
nuclear radio emission at 20 cm and 1.4 GHz, respectively. The largest
contribution to this SN rate comes from A1. These SN explosions would power a
galactic superwind which manifests itself as a gas outflow in the H$\alpha$
and Br$\gamma$ images (cf. Heckman et al. 1987, Paper I).

The predicted mechanical energy flux released by the stellar winds and SN
explosions in A1, A2 and B1 is $\simeq$ 6 $\times$ 10$^{43}$ ergs s$^{-1}$
from both the continuous and instantaneous star-formation scenarios. This
value is in agreement with the predictions of Heckman et al. (1990). Indeed,
if a total IR luminosity of $(3 - 11) \times 10^{11}$ L$_{\odot}$ is assumed
for NGC~6240, equation (11c) in Heckman et al. gives an energy flux from SN
explosions and stellar winds in the range $(2 - 8) \times 10^{43}$ ergs
s$^{-1}$. According to the superbubble model by Mac Low \& McCray (1988), a
conversion factor of total wind/SN energy flux to kinetic energy of the gas of
$\sim$ 30$\%$ implies a kinetic energy of $\leq$ 2 $\times$ 10$^{43}$ ergs
s$^{-1}$. Adopting a velocity spread of the H$\alpha$ filaments of 1000 km
s$^{-1}$ (Heckman et al. 1990), we have computed a galactic wind, mass-loss
rate of about 44 M$_{\odot}$ yr$^{-1}$.

How do the STARBURST99 models compare with the observed, soft X-ray emission
in NGC~6240? Schulz et al. (1998) determined, from ROSAT observations, a soft
X-ray luminosity of about 10$^{42}$ ergs s$^{-1}$. Based on the superwind
model by Mac Low \& McCray (1988) and Heckman et al. (1996), Schulz et al.
suggested that this luminosity is due to an input energy flux of 10$^{44}$
ergs s$^{-1}$, equivalent to 3 SNe per year. Such a flux can expand a
superbubble out to a radius of 10 kpc in 3 $\times$ 10$^7$ yr. The
calculations assumed an ISM density of 0.1 cm$^{-3}$ and a filling factor of
the X-ray emitting gas of about 0.2. We have repeated the exercise of Schulz
et al. by assuming an input energy of 6 $\times$ 10$^{43}$ ergs s$^{-1}$ (from
the STARBURST99 fits to the observed photometry of A1, A2 and B1 with
Z = Z$_{\odot}$) and a
kinetic energy of 2 $\times$ 10$^{43}$ ergs s$^{-1}$. From Schulz et al.'s
equation (3), we estimate a soft X-ray luminosity of $\simeq$ 6 $\times$
10$^{41}$ ergs s$^{-1}$, which agrees with the observed X-ray luminosity
within a factor of $<$ 2. The agreement strongly depends on the uncertainties
in the observations and the model assumptions. However, it could be improved
if, for example, the conversion factor of input energy flux to kinetic energy
were higher than 27$\%$ (up to perhaps 50$\%$) and/or the filling factor of
the superbubble were smaller than 0.2, by perhaps a factor of 2.

\subsection{Star Formation Lifetimes and Histories}

The nuclear condensations A1, A2 and B1 appear to be ``embedded'' in a
circumnuclear population of $\sim$1 Gyr old (Genzel et al. 1998, Tecza et al.
2000), possibly implying that the NGC~6240 nuclei have been undergoing
a large spread in the star formation activity. 
This co-existence of old and young stellar populations has also been
seen in the Antennae nuclei, where Berta et al. (2003) have been able to date
two star-formation episodes; one occurred about 1 Gyr ago and the other is
still ongoing. The authors suggest that the bimodal star-formation history in
the Antennae nuclei is the result of a double encounter in the dynamical
evolution of the galaxies' merger. Similar star formation patterns are seen in
other classes of interacting galaxies; e.g. the Seyfert 1 NGC~3227 (Schinnerer
et al. 2001). However, it is also possible that the pre-merger systems could
provide a substantial fraction of the older stars seen in NGC~6240. The
correlation between nuclear black hole mass and the stellar mass of the bulges
of non-interacting galaxies implies that the pre-merger galaxies in NGC~6240
both had significant bulges (Kormendy \& Richstone 1995, Marconi \& Hunt
2003).

The data also indicate that substantial variations in SFRs occur over very
short time spans, $\leq$ 10 Myr, in mergers. For example, in Arp~299
Alonso-Herrero et al. (2000) detected an extended period of star formation
that began $>$ 7 Myr before the most recent episodes with ages of $\sim$ 4
Myr. The extremely young stellar population ages of $\simeq$ 4 Myr that we
find for the three NGC~6240 nuclei is another indication of this effect. Rapid
variations in SFRs might be expected to be associated with the compact, high
density central gas concentrations that are characteristic of ULIRGs and
related classes of starburst galaxies (e.g. Solomon et al. 1997). These
circumnuclear regions have sizes of $0.1 - 0.5$ kpc and internal velocities of
$\geq$ 100 km s$^{-1}$ and associated dynamical scales of only $\sim$ 10$^7$
yr. It should be no surprise that star formation near the centres of merging
galaxies is dynamic, but given the huge gas concentrations in these regions,
typically $\sim$ 10$^9$ M$_{\odot}$, the effects of SFR fluctuations are
spectacular. Given that our mass estimates for A1, A2 and B1 are comparable to
within a factor of $< 2$ between the instantaneous and continuous star
formation scenarios, our results are fairly robust with respect to the star
formation history adopted.

The time to produce the stellar mass in a system if the current SFR remains
constant is an useful indicator of evolutionary rates. From Table 4 we see
that this scale is $\tau_* \sim$ 10$^7$ yr for all the three nuclear
condensations. Another estimate of the expected time scales for the evolution
of the NGC~6240 nuclei can, in principle, be obtained by examining the mass
balance of the nuclear interstellar medium, 
$$\dot M_{\rm ISM} = \dot M_{\rm a} - {\rm SFR} - \dot M_{\rm wind},$$ 
where $\dot M_{\rm a}$ is the gas accretion rate into each nucleus and $\dot
M_{\rm wind}$ is the mass loss rate out of each nuclear region fueling the
galactic wind. Using nucleus A1 as an example, we have $\dot M_{\rm wind}
\leq$ 40 M$_{\odot}$ yr$^{-1}$ (cf. Sect. 4.1.3 and Heckman et al. 1990) and
SFR $\sim$ 270 M$_{\odot}$ yr$^{-1}$. The major unknown is the accretion rate
into the nucleus. We can estimate this rate by $\dot M_{\rm a} \sim M_{\rm
g,nuclear}/t_{\rm cross,nuclear}$ where $t_{\rm cross,nuclear} = R/V \sim$
10$^7$ yr. Adopting a nuclear gas mass of 10$^9$ M$_{\odot}$ from Tacconi et
al. (1999), we then estimate $\dot M_{\rm a} \sim$ 100 M$_{\odot}$ yr$^{-1}$.

The characteristic time scale for the starburst in nucleus A1 is then
$\tau_{\rm burst} \sim - M_g$ / $\dot M_a <$ 10$^7$ yr. The combination of
high SFRs, large flow velocities, and small sizes of the relevant regions
demand rapid variations in SFRs. Even if one invokes a maximal estimate of the
{\it total} mass of interstellar gas in NGC~6240 of 3 $\times$ 10$^{10}$
M$_{\odot}$ (Georgakakis et al. 2000, allowing for the mass of $^4$He) and a
short dynamical time scale of 100 Myr for the entire system, we see that the
current starburst can be sustained for $<$ 100 Myr. The actual lifetime of a
burst with the current SFR is probably considerably shorter under more
realistic assumptions as to how much gas actually can fuel the nuclear
starburst region.

\subsection{Evolution of the Nuclei}
\subsubsection{Compact Disks and the AGN}

While the two AGN in NGC~6240 and their associated star clusters appear to be
simple objects at our angular resolution, kinematic studies supply a glimpse
of their true complexity. Observations of the central stellar velocity field
by Tecza et al. (2000) from the ground in the K-band show pronounced and
organized radial velocity gradients around nuclei A1 and B1. They interpret
these velocity patterns as resulting from the rotation of inclined stellar
objects, such as disks or flattened bulges, and by applying this model derive
dynamical masses of $\ge$10$^9$~M$_{\odot}$. High angular resolution
observations of emission line velocities and widths by Gerssen et al. (2003)
show more complexity. Their results could arise from some combination of
nuclear disks and large-scale gas flows associated with the starbursts and
ongoing interaction. Disturbed large-scale flows in the gas that are largely
decoupled from the nuclei are also consistent with what is seen in CO emission
(Tacconi et al. 1999), H$_2$ IR emission lines (Ohyama et al. 2000, Ohyama
et al. 2003) and in HI absorption against the AGNs (Beswick et al.
2001).

Despite the complex gas kinematics seen towards A1 and B1, it is likely that
both AGNs are surrounded by compact gas disks that have substantial rotational
support. This model is consistent with the K-band stellar kinematics, and also
with the presence of OH maser emission from B1 (Hagiwara, Diamond, \& Miyoshi
2003). Nuclear maser emission in ULIRGs is thought to occur in rotating disks
of dense gas with sizes of a few hundred pc, which provide the large column
densities of gas at near constant velocity that are required for masers (cf.
Philstr\"om et al. 2001). When other nuclear starbursts with AGNs are studied
with sufficient precision, molecular gas disks with radii of $\sim$100~pc and
masses $>$10$^8$~M$_{\odot}$ are found (e.g., Mrk 231; Bryant \& Scoville
1996, Arp~220; Scoville et al. 1997, NGC~4303; Colina \& Arribas 1999). While
such structures seem to be generic features of strong interactions involving
gas-rich galaxies, it is not clear which physical properties determine their
size and mass scales (e.g., Downes \& Solomon 1998, Jogee et al. 2002).

\subsubsection{AGN and Pre-Merger Bulge Masses}

The combined total luminosity of the two AGNs in NGC~6240 is estimated by Lira
et al. (2002) to be $\sim$5 $\times$ 10$^{45}$~erg~s$^{-1}$, comparable to
what is produced by the starburst (see also Lutz et al. 2003). We therefore
assume that each nucleus produces $\sim$10$^{12}$~L$_{\odot}$. A lower bound
estimate for the black hole masses of $M_{\rm BH} \sim $10$^{7.5}$~M$_{\odot}$
then follows from the Eddington limit argument. Assuming that most of this
mass was deposited in the black holes before the present merger, we can
estimate properties of the pre-merger system bulges on the basis of empirical
correlations between $M_{\rm BH}$ and host galaxy bulge properties. Thus we
find that the pre-merger bulge velocity dispersion would have been
$\approx$150~km~s$^{-1}$ and the bulge B-band luminosities about
10$^{9.5}$~L$_{\odot}$ corresponding to a stellar mass near
10$^{10}$~M$_{\odot}$. The presence of dual AGN in NGC~6240 is consistent with
it being a major merger involving two giant spiral galaxies.

The similarity between conditions in Arp~299 and those in NGC~6240 suggests
another possible evolutionary trend. NGC~6240 and Arp~299 are the only two merging
systems known to contain dual AGNs. In both cases, it appears that the amount
of mass going into the circumnuclear starbursts considerably exceeds that
being accreted by the nuclear black holes. Since energy production from
accretion onto a black hole is more than order of magnitude more efficient
than what is provided by nuclear energy in stars, and the AGN and starburst
luminosities are comparable in NGC~6240, we predict a black hole accretion
rate of $\sim3-30$~M$_{\odot}$~yr$^{-1}$ in nucleus A1. From our earlier
discussion of the lifetime of the starburst, we take 30~Myr as a generous
estimate of the main starburst lifetime. We then see that if the system
continues in its present state, the black holes could grow in mass by
$\ge$10$^8$~M$_{\odot}$ prior to the eventual merger of the binary nuclei into
a single object. Our analysis thus suggests the nucleus of the final NGC~6240
merger will have $M_{\rm BH} \ge $2$ \times $10$^8$~M$_{\odot}$.

\subsubsection{The A2 Cluster}

Our photometric study demonstrates that nucleus A2 is also a huge
concentration of young stars. However, it shows no evidence for being an AGN,
and therefore could be an example of an extraordinarily luminous star cluster.
For example, young super star clusters with similar masses are found in the
centre of the Arp~220 merger, where the extraordinary N2 and N3 clusters
described by Shioya, Taniguchi, \& Trentham (2001) are not necessarily
associated with nuclei. The fate of such clusters is somewhat uncertain.
However, due to their high masses, these clusters likely will have short
lifetimes driven by the dynamical friction time scale for them to spiral into
the centres of their host galaxies, where they will merge with the
AGNs/nuclear star clusters. This type of event can help to populate the bulge
of the remnant galaxy (cf. Noguchi 1999, Kim, Morris, \& Lee 1999).

\subsubsection{The Fate of NGC~6240}

The B1, A1 and A2 integrated masses give a mass density of about 16.0, 1.2
$\times$ 10$^3$ and 19.0 M$_{\odot}$ pc$^{-3}$, respectively, as measured in a
circular aperture of radius 0$''$.1, and assuming continuous star formation.
For a single and instantaneous burst of star formation, the cluster mass
density increases to 67.0, 1.4 $\times$ 10$^3$ and 80.0 M$_{\odot}$ pc$^{-3}$.
In either scenario, these values are similar to the mass density of the
molecular gas ($10^2 - 10^3$ M$_{\odot}$ pc$^{-3}$) integrated over the
central R $\le 0.5 - 1$ kpc of most nearby ULIRGs (Sanders \& Mirabel 1996),
where 0.5 kpc translates into 1$''$ at the distance of NGC 6240, i.e. the
entire double core of NGC 6240. Evans et al. (2002) estimated the core mass
density in molecular gas for a sample of ULIRGS with double nuclei and obtained
values of 970 to 8.6 $\times$ 10$^3$ M$_{\odot}$ pc$^{-3}$ per nucleus.
Compared with this sample, A1 is similar to IRAS 12112$+$0305SW and IRAS
14348$+$1447SW. In addition, the mass density of A1 falls in the low-mass
density tail of the distribution of early-type galaxies, whose masses were
measured in an aperture with a radius of 0$''$.1 by Faber et al. (1997).
Specifically, A1 appears to fit in with the ``core'' galaxies of Faber et
al.'s sample. Faber et al. (1997) separate early-type galaxies into two
categories: the power-law galaxies with a fairly steep surface profile in
their surface brightness, extending inward to the smallest resolvable radius,
and the core galaxies showing a broken power-law profile which changes to a
significantly shallower slope at a specific break radius. Core galaxies are
bright objects with M$_V \leq -20.5$ (H$_0$ = 80 km s$^{-1}$ Mpc$^{-1}$) and
are found in clusters as well as in the field.

The mass densities derived for B1, A1 and A2 suggest that, whatever the nature
of the burst is, if their fate is to merge together, then B1, A1 and A2 would
likely form the core of an elliptical or at least a large bulge of an
early-type spiral galaxy, which NGC 6240 could evolve into, once the merger of
the parent galaxies is completed. The findings of Komossa et al. (2003)
support this scenario. The presence of a heavily obscured AGN in the A nucleus
has long been known, but the definitive detection of an AGN in the B nucleus,
hence of supermassive binary black holes, comes from the very recent Chandra
ACIS-S data. According to Milosavljevi\'c \& Merritt (2001) and
Milosavljevi\'c et al. (2002), during a galactic merger the two black holes
fall to the centre of the merger and form a bound pair. Their spiraling
towards each other releases their binding energy to the surrounding stars,
modifying the central stellar density of the merger. Consequently, the initial
central cusp in the stellar density evolves into a shallower distribution,
$\rho \sim r^{-1}$, similar to the luminosity profile observed at the centre
of elliptical core galaxies. This evolution occurs within $10^6 - 10^7$ yr
after the binary black hole has formed; the cusp continues to flatten
thereafter as the binary ejects stars via gravitational kicks away from the
nucleus of the merger. The process should terminate when the binary reaches a
separation of between 0.01 and 1 pc.

The black holes detected in the double nucleus of NGC 6240 are about 1.5 arcsec
apart (Komossa et al. 2003), which translates into 700~pc at a distance of 98
Mpc, therefore NGC 6240 may be at the beginning of the ``cusp coalescence''
defined by Milosavljevi\'c \& Merritt (2001) and may transform into an
elliptical core galaxy. This agrees with the conclusion by Genzel et al.
(2001) who measured a rotational velocity and a velocity dispersion in
NGC~6240 consistent with the properties of elliptical and lenticular systems.

\section{Summary}

We have analysed HST data available for three discrete sources identified in
the double nucleus of NGC~6240. We have fitted the observed colours to
STARBURST99 models to estimate their ages, masses and intrinsic reddenings.
Our fitting technique (cf. Paper I) relies on observed colours corrected for
the Galactic extinction in the direction of NGC~6240 and develops through five
steps: {\it i)} the synthetic colours at each time mesh-point of the model are
reddened by E(B-V) increasing from 0.0 to 5.0 mag in steps of 0.01 mag; {\it
ii)} at each step in E(B-V), a $\chi_{\rm tot}^2$ is computed as the sum of
the colour differences between observations and evolutionary tracks weighted
by the observational errors and squared; {\it iii)} for each time mesh-point
of the model, a minimum $\chi_{\rm tot}^2$ is extracted from all the
$\chi_{\rm tot}^2$ values obtained by varying the reddening and the model age,
magnitudes and colours; {\it iv)} once all the time mesh-points of the model
have been fitted, the smallest $\chi_{\rm tot}^2$ 
($\chi_{\rm MIN}^2$) is identified among all
$\chi_{\rm tot}^2$ minima and its corresponding E(B-V), model age, magnitudes
and colours are assigned to the cluster concerned; {\it v)} the selected model
magnitudes are reddened by the final selected E(B-V), corrected for the
distance modulus to NGC~6240 and scaled to the observed apparent magnitudes.
The assumed SFR is multiplied by this magnitude scaling factor to derive the
actual SFR of the sources. We have made use of two sets of STARBURST99
models, computed for two different metallicities, solar- and LMC-like.
The results obtained from these two sets of evolutionary tracks agree
reasonably well and are within the uncertainties of our fitting procedure.
Therefore, the subsequent analysis and discussion of the properties of
the nuclear sources have been based on the STARBURST99 models with solar
metallicity. The nuclear condensations in NGC~6240 have four
independent colours available, (B-V), (V-I), (V-J) and (V-K), which allow us
to determine a unique combination of age, reddening and SFR with relatively
small uncertainties.

The probability distributions for the nuclear sources in NGC~6240 have been
derived from the fits with $\chi_{\rm tot}^2 \le$ 1.5 times the 
$\chi_{\rm min}^2$ value of the fitting procedure. Under the
assumption that these sources are clusters, they are dated as young as 4 Myr
and are heavily obscured, with E(B-V) = $1 - 2.3$ or A$_V = 4 - 9$ in
agreement with spectroscopic measurements. Their mass estimates, however, are
extremely high, from 3 $\times$ 10$^7$ to 10$^9$ M$_{\odot}$. A mass of 10$^7$
M$_{\odot}$, as found for B1, has been so far determined and confirmed
spectroscopically only for cluster W3 in NGC 7252, while a lower mass of a few
10$^5$ M$_{\odot}$ has been estimated for the nuclear clusters in NGC 4303 and
NGC 4449. On the other hand, the tidal radius of the nuclear sources in
NGC~6240 is larger than measured for bona-fide compact clusters and thus this
favours the earlier suggestion that these sources are giant H{\sc ii} regions
instead. In this case, a STARBURST99 model with continuous star formation is
probably more appropriate, which results in:
\par\noindent 
1) young ages of about 4.3 $\times$ 10$^6$ yr and reddenings in the range 0.8
$\le$ E(B-V) $\le$ 2.4 mag;
\par\noindent
2) SFRs between 1.5 (B1) and 266 (A1) M$_{\odot}$ yr$^{-1}$,
which define integrated masses from 7 $\times$ 10$^6$ to 10$^9$ M$_{\odot}$;
\par\noindent
3) mass densities from 16 (B1) to 1200 (A1) M$_{\odot}$ pc$^{-3}$.
\par\noindent
4) a SN rate of 1 SN yr$^{-1}$ and a galactic wind of 44 M$_{\odot}$
yr$^{-1}$. Comparison with previously determined SN rates and galactic wind
estimates based on NGC~6240's L$_{\rm IR}$ indicates that the nuclear sources
dominate the galaxy's total IR luminosity and star formation activity.
\par
Irrespective of the scenario assumed, the properties of the nuclear sources
are such to likely sustain the observed soft X-ray emission. They can also
be used to estimate a mass of $\sim$ 10$^{10}$ M$_{\odot}$ for the bulge
of the parent galaxies and a black-hole mass of $\geq$ 2 $\times$ 10$^8$
M$_{\odot}$ for the end-product of the coalescence of the binary black hole.
\par
Star formation rates such as those found for B1 and A2 are commonly observed
in nearby blue compact galaxies such as NGC 7673 and Lyman break galaxies,
while the SFR value derived for A2 is typical of warm ULIRG galaxies with
double nuclei. Similarly, the mass density in A1 is consistent with the mass
densities of molecular gas in ULIRGs and the stellar mass densities measured in
elliptical core galaxies. The supermassive binary black holes recently
discovered in NGC~6240 (Komossa et al. 2002) could in time transfer energy to
the stars in A1, A2 and B1, thus lowering the central stellar density in
NGC~6240 nuclei to a flat $\rho \sim r^{-1}$ profile, typical of elliptical
core galaxies (cf. Milosavljevi\'c \& Merritt, 2001). The balance between the
relative efficiencies of star formation, dynamical friction, and the slingshot
effect of the coalescing binary black hole will ultimately determine whether
the merger in NGC~6240 will produce a cusp or a core elliptical.

\acknowledgements
It is a pleasure to thank Claus Leitherer whose valuable comments
improved the paper.
AP would like to thank F. van den Bosch and M. Carollo for stimulating
discussions. JSG thanks the University of Wisconsin Graduate School for
partial support of this research and the European Southern Observatory for its
hospitality while working on this project. This paper is based on observations
made with the NASA/ESA Hubble Space Telescope, obtained from the data archive
at the Space Telescope Science Institute (STScI). STScI is operated by the
association of Universities for Research in Astronomy, Inc. under NASA
contract NAS 5-26555. This research has made extensive use of NASA's
Astrophysics Data System Abstract Service.


\begin{thebibliography}{}

\bibitem{} Alonso-Herrero, A., Rieke, G.H., Rieke, M.J., Scoville, N.Z., 2000, ApJ, 532, 845
\bibitem{} Anders, P., Fritze-von Alvensleben, U., 2003, A\&A, 401, 1063
\bibitem{} Ballo, L., Braito, V., Della Ceca, R., Maraschi, L., Tavecchio, F., Dadina, M.,
2003, MNRAS, 335, 1176 
\bibitem{} Barbieri, C., Rafanelli, P., Schulz, H. et al. 1993, A\&A, 273, 1
\bibitem{} Berta, S., Fritz, J., Franceschini, A., Bressan, A., Pernechele, C., 2003, A\&A, 403, 119 
\bibitem{} Beswick, R.J., Pedlar, A., Mundell, C.G., Gallimore, J.F., 2001, MNRAS, 325, 151
\bibitem{} B\"oker, T., van der Marel, R.P., Mazzuca, L., Rix, H.-W., Rudnick, G., Ho, L.C., Shields, J.C., 2001, AJ, 121, 1473
\bibitem{} B\"oker, T., Laine, S., van der Marel, R.P., Sarzi, M., Rix, H.-W., Ho, L.C., Shields, J.C.,, 2002, AJ, 123, 1389
\bibitem{} Bryant, P.M., Scoville, N.Z., 1999, AJ, 117, 2632
\bibitem{} Bushouse, H.A., Borne, K.D., Colina, L., Lucas, R.A., Rowan-Robinson, M., Baker, A.C., Clements, D.L., Lawrence, A., Oliver, S., 2002, ApJS, 138, 1
\bibitem{} Calzetti, D., 2001, PASP, 113, 1449
\bibitem{} Carollo, C.M., Stiavelli, M., de Zeeuw, P.T., Mack, J., 1997, AJ, 114, 2366
\bibitem{} Carollo, C.M., Stiavelli, M., de Zeeuw, P.T., Seigar, M., Dejonghe, H., 2001, ApJ, 546, 216 
\bibitem{} Cervi\~no, M., Valls-Gabaud, D., Luridiana, V., Mas-Hesse, J.M., 2002, A\&A, 381, 51
\bibitem{} Cohen, J.G., Freeman, K.C., 1991, AJ, 101, 483
\bibitem{} Colbert E.J.M., Wilson A.S., Bland-Hawthorn J., 1994, ApJ, 436, 89
\bibitem{} Colina, L., Arribas, S. 1999, ApJ, 514, 637
\bibitem{} Colina, L., Gonz\'alez Delgado, R., Mas-Hesse, J.M., Leitherer, C., 2002, ApJ, 579, 545 
\bibitem{} Cui, J., Xia, X.-Y., Deng, Z.-G., Mao, S., Zou, Z.-L., 2001, AJ, 122, 63 
\bibitem{} de Grijs, R., Bastian, N., Lamers, H.J.G.L.M., 2003a, ApJ, 583, L17
\bibitem{} de Grijs, R., Anders, P., Bastian, N., Lynds, R., Lamers, H.J.G.L.M., O'Neil Jr., E.J., 2003b, MNRAS, 343, 1285
\bibitem{} de Grijs, R., O'Connell, R.W., Gallagher, J.S., 2001, AJ, 121, 768
\bibitem{} DePoy D.L., Becklin E.E., Wynn-Williams C.G., 1986, ApJ, 307, 116
\bibitem{} Djorgovski, S.G., 1993, in Structure and Dynamics of Globular Clusters, eds. S.G. 
Djorgovski \& G. Meylan (San Francisco: ASP), p. 373
\bibitem{} Dowens, D., Solomon, P.M., 1998, ApJ, 507, 615
\bibitem{} Evans, A.S.,  Frayer, D.T., Surace, J.A., Sanders, D.B., 2001, AJ, 121, 3286
\bibitem{} Evans, A.S., Mazzarella, J.M., Surace, J.A., Sanders, D.B., 2002, ApJ, 580, 749   
\bibitem{} Faber, S.M., Tremaine, S., Ajhar, E.A., Byun, Y.-I., Dressler, A., Gebhardt, K., Grillmair, K., Kormendy, J., 
Lauer, T.R., Richstone, D., 1997, AJ, 114, 1771
\bibitem{} Farrah, D., Rowan-Robinson, M., Oliver, S., Serjeant, S., Borne, K., Lawrence, A., Lucas, R.A., Bushouse, H., Colina, 
L., 2001, MNRAS, 326, 1333 
\bibitem{} Fosbury R.A.E., Wall J.V., 1979, MNRAS, 189, 79
\bibitem{} Fritze-von Alvensleben, U., 1998, A\&A, 336, 83
\bibitem{} Fritze-von Alvensleben, U., 1999, A\&A, 342, L25
\bibitem{} Georgakakis, A., Forbes, D.A., Norris, R.P., 2000, MNRAS, 318, 124
\bibitem{} Genzel, R., Lutz, D., Sturm, E., et al., 1998, ApJ, 498, 579
\bibitem{} Genzel, R., Tacconi, L.J., Rigopoulou, D., Lutz, D., Tecza, M., 2001, ApJ, 563, 527
\bibitem{} Gerssen, J., van der Marel, R. P., Axon, D. J., Mihos, C.,
   Hernquist, L., \& Barnes, J. E. 2001, in The Central Kiloparsec of
   Starbursts and AGN: The La Palma Connection, ASP Conference
   Proceedings Vol. 249. eds. J. H. Knapen, J. E. Beckman, I. Shlosman, \&
   T. J. Mahoney, p.~665 (San Francisco: Astronomical Society of
   the Pacific)
\bibitem{} Gerssen, J., van der Marel, R.P., Axon, D., Mihos, C., 
Hernquist, L., Barnes, J., 2003, AJ, submitted (astro-ph/0310029)
\bibitem{} Gonz\'alez Delgado, R.M., Heckman, T., Leitherer, C., 2001, ApJ, 546, 845
\bibitem{} Gonz\'alez Delgado, R.M., Heckman, T., Leitherer, C., Meurer, G., Krolik, J., Wilson, A.S.,
Kinney, A., Koratkar, A., 1998, ApJ, 505, 174
\bibitem{} Goudfrooij, P., Victoria Alonso, M., Maraston, C., Minniti, D., 2001, MNRAS, 328, 237
\bibitem{} Grillmair, C.J., Ajhar, E.A., Faber, S.M., Baum, W.A., Holtzman, J.A., Lauer, T.R., Lynds, C.R., O'Neil Jr., E.J., 1996, AJ, 111, 2293 
\bibitem{} Hagiwara, Y., Diamond, P.J., Miyoshi, M., 2003, A\&A, 400, 457 
\bibitem{} Heckman, M.T., Armus, L., Miley, G.K., 1987, AJ, 92, 276
\bibitem{} Heckman, M.T., Armus, L., Miley, G.K., 1990, ApJS, 74, 833
\bibitem{} Heckman, M.T., Dahlem, M., Eales, S.A., Fabbiano, G., Weaver, K., 1996, ApJ, 457, 616
\bibitem{} Holtzman, J.A., Burrows, C.J., Casertano, S., Hester, J.J., Trauger, J.T., Watson, A.M.,
Worthey, G., 1995, PASP, 107, 1065
\bibitem{} Iwasawa, K., Comastri, A., 1998, MNRAS, 297, 1219
\bibitem{} Jogee, S., Baker, A. J., Sakamoto, K., Scoville, N. Z.,
Kenney, J. D. P. 2001 in The Central kpc of Starbursts and AGN,
ASP Conf. Ser., Vol. 249, eds. J. H. Knapen, J. E. Beckman,
I. Shlosman, T. J. Mahoney, p. 612
\bibitem{} Kim, S. S., Morris, M., Lee, H. M., 1999, ApJ, 525, 228
\bibitem{} Komossa, St., Burwitz, V., Hasinger, G., Predehl, P., Kaastra, J.S., Ikebe, Y., 2003,
ApJ, 582, L15
\bibitem{} Kormendy, J., Richstone, D., 1995, ARA\&A, 33, 581
\bibitem{} Lan\c{c}on, A., Mouhcine, M., 2000, in Massive Stellar Clusters, ASP Conf.
er. 211, eds. A. Lan\c{c}on \& . M. Boily, p. 34
\bibitem{} Leitherer, C., Schaerer, D., Goldader, J.D., et al., 1999, ApJS, 123, 3
\bibitem{} Lira, P., Ward, M.J., Zezas, A., Murray, S.S., 2002, MNRAS, 333, 709 
\bibitem{} Lutz, D., Sturn, E., Genzel, R., Spoon, H.W.W., Moorwood, A.F.M., Netzer, H.,
Sternberg, A., 2003, A\&A, 409, 867 
\bibitem{} Mac Low, M.-M., McCray, R., 1988, ApJ, 324, 776
\bibitem{} Maraston, C., Kissler-Patig, M., Brodie, J.P., Barmby, P., Huchra, J.P., 
2001, A\&A, 370, 176
\bibitem{} Marconi, A., Hunt, L.K., 2003, ApJ, 589, L21
\bibitem{} Meurer, G.R., 1995, Nature, 375, 742
\bibitem{} McCall, M.L., Rybski, P.M., Shileds, G.A., 1985, ApJS, 57, 1
\bibitem{} Milosavljevi\'c, M., Merritt, D., 2001, ApJ, 563, 34
\bibitem{} Milosavljevi\'c, M., Merritt, D., Rest, A., van den Bosch, F.C., 2002, MNRAS, 331, L51
\bibitem{} Neugebauer, G., Soifer, B.T., Beichman, C.A., Aumann, H.H., Chester, T.J.,
Gautier, T.N., Lonsdale, C.J., Gillett, F.C., Hauser, M.G., Houck, J.R., 1984, Science,
224, 14
\bibitem{} Noguchi, M. 1999, ApJ, 514, 77
\bibitem{} Ohyama Y., et al., 2000, PASJ, 52, 563 
\bibitem{} Ohyama, Y., Yoshida, M., Takata, T. 2003, AJ, in press (astro-ph/0307517)
\bibitem{} Pasquali, A., de Grijs, R., Gallagher, J.S., 2003, MNRAS, 345, 161 (Paper I)
\bibitem{} Pettini, M., Shapley, A.E., Steidel, C.C., Cuby, J.-G., Dickinson, M., Moorwood, A.F.M., Adelberger, K.L., 
Giavalisco, M., 2001, ApJ, 554, 981 
\bibitem{} Philstr\"om, Y. M., Conway, J. E., Booth, R. S., Diamond,
P. J., Polatidis, A G. 2001, A\&A, 377, 413
\bibitem{} Pisano, D., Kobulnicky, H., Guzm\'an, R., Gallago, J., Bershady, M., 2001, AJ, 122, 1194
\bibitem{} Rafanelli, P., Schulz, H., Barbieri, C., Komossa, S., Mebold, U., Baruffolo, A., Radovich, M., 1997, A\&A, 327, 901
\bibitem{} Rieke, G.H., Cutri, R.M., Black, J.H., Kailey, W.F., McAlaray,
C.W., Lebofsky, M.J., Elston, R., 1985, ApJ, 290, 116
\bibitem{} Sanders, D.B., 2001, IAU Coll. 184, eds. R.F. Green, E.Ye. Khachikian \& D.B.
Sanders 
\bibitem{} Sanders, D.B., Mirabel, I.F., 1996, ARA\&A, 34, 749
\bibitem{} Schinnerer, E., Eckart, A., Tacconi, L.J., 2001, ApJ, 549, 254
\bibitem{} Schlegel, D.J., Finkbeiner, D.P., Marc, D., 1998, ApJ, 500, 525
\bibitem{} Schulz, H., Komossa, S., Bergh\"ofer, T.W., Boer, B., 1998, A\&A, 330, 823
\bibitem{} Schweizer, F., Seitzer, P., 1998, AJ, 116, 2206
\bibitem{} Scoville, N.Z., Evans, A.S., Thompson, R., Rieke M., Hines, D.C., Low, F.J., Dinshaw, N., Surace, J.A., Armus, L., 2000, AJ, 119, 991
\bibitem{} Simpson, C., Forbes, D.A., Baker, A.C., Ward, M.J., 1996, MNRAS, 189,
7904
\bibitem{} Shioya, Y., Taniguchi, Y., Trentham, N., 2001, MNRAS, 321, 11
\bibitem{} Soifer, B.T., Neugebauer, G., Helou, G., Lonsdale, C.J., Hacking, P.,
Rice, W., Houck, J.R., Low, F.J., Rowan-Robinson, M., 1984, ApJ, 283, L1
\bibitem{} Solomon, P.M., Dowens, D., Radford, S.J.E., Barrett, J.W., 1997, ApJ, 478, 144 
\bibitem{} Surace, J.A., Sanders, D.B., Evans, A.S., 2000, ApJ, 529, 170
\bibitem{} Surace, J.A., Sanders, D.B., Vacca, W.D., Veilleux, S., Mazzarella, J.M., 1998, ApJ, 492, 116
\bibitem{} Tacconi, L.J., Genzel, R., Tecza, M., Gallimore, J.F., Downes,
D., Scoville N.Z., 1999, ApJ, 524, 732
\bibitem{} Tecza, M., Genzel, R., Tacconi, L.J., Anders, S., Tacconi-Garman, L.E., Thatte, N., 2000, ApJ, 537, 178
\bibitem{} Toomre, A., 1977, in The Evolution of Galaxies and Stellar Populations, eds. B. Tinsley \& R. Larson (New Haven: Yale Univ. Press), p. 401
\bibitem{} Veilleux, S., Kim, D.-C., Sanders, D.B., 1999, ApJ, 522, 113
\bibitem{} Veilleux, S., Kim, D.-C., Sanders, D.B., Mazzarella, J.M., Soifer, B.T., 1995, ApJS, 98, 171
\bibitem{} Vignati, P. et al., 1999, A\&A, 349, L57
\bibitem{} van der Werf, P.P., Genzel, R., Krabbe, A. et al., 1993, ApJ, 405, 522
\bibitem{} Whitmore, B.C., Heyer, I., 1998, ISR WFPC2 97-08
\bibitem{} Whitmore, B.C., Schweizer, F., Leitherer, C., Borne, K., Robert, C., 1993, AJ, 106, 1354
\bibitem{} Whitmore, B.C., Zhang, Q., Leitherer, C., Fall, M.S., Schweizer, F., Miller, B.W., 1999,
AJ, 118, 1551
\bibitem{} Wright, G.S., Joseph, R.D., Meikle, W.P.S., 1984, Nature, 309, 430
\bibitem{} Zepf, S.E., Ashman, K.M., 1993, MNRAS, 264, 611
\bibitem{} Zwicky, F., Herzog, E., Wild, P., 1961, Catalogue of Galaxies
and of Clusters of Galaxies (Pasadena: Caltech), vol. 1

\end{thebibliography}
\end{document}